\documentstyle[aps,psfig]{revtex}
\begin{document}
\draft
\title{Quark Nuggets as Baryonic Dark Matter}
\author{Jan-e Alam} 
\address{Variable Energy Cyclotron Centre, 1/AF Bidhan Nagar, Calcutta
700 064, INDIA.}
\author{Sibaji Raha}
\address{Bose Institute, 93/1 A. P. C. Road, Calcutta 700 009, INDIA.}
\author{Bikash Sinha}
\address{Variable Energy Cyclotron Centre, 1/AF Bidhan Nagar, Calcutta
700 064, INDIA.\\
Saha Institute of Nuclear Physics, 1/AF Bidhan Nagar, Calcutta
700 064, INDIA.}
\date{\today}
\parindent=20pt
\maketitle
\vskip 0.4in
\centerline{\bf{\normalsize Abstract}}
\vskip 0.15in
The cosmic first order phase transition from quarks to hadrons, occurring
a few microseconds after the Big Bang, would lead to the formation of
quark nuggets which would be stable on a cosmological time scale, if the
associated baryon number is larger than a critical value. We examine the
possibility that these surviving quark nuggets may not only be viable
candidates for cold dark matter but even close the universe.

\vskip 0.4in
\noindent

According to the wisdom of the {\it standard model}, the universe underwent,
a few microseconds after the big bang, a phase transition from quarks to
hadrons. The cosmological implications of this, {\it presumably first
order}, phase transition would be far-reaching. Schramm and collaborators
\cite{schramm2,schramm3} (see also \cite{vanhove}) argued that
the associated fluctuations may lead to the formation of primordial black
holes, which could be as large as $M_\odot$, the solar mass, for
fluctuations around the horizon scale. They recently suggested
\cite{schramm2} that these black holes could even be the candidates
for the Massive Compact Halo Objects (MACHO's) \cite{macho1,macho2}, which
had of late been discovered in the halo of the Milky way, in the direction
of the Large Magellanic Cloud (LMC) through their gravitational microlensing
properties. On the other hand, a first order phase transition scenario
involving bubble nucleation at a critical temperature $T_c\sim 100-200$ MeV
should lead to the formation of quark nuggets (QN) \cite{witten}, made of
$u$, $d$ and $s$ quarks at a density $\geq$ nuclear density. If these
primordial QN's existed till the present epoch, they could be possible
candidates for the dark matter \cite{witten}. Such a possibility would be
aesthetically rather pleasing, as it would not require any exotic physics
nor would the success of the primordial (Big Bang) nucleosynthesis scenario
be affected \cite{yang,schaeffer,rana,madsen}.

The central question in this context then is whether the primordial QN's can
be stable on a cosmological time scale. In a recent work \cite{pijush}
using the chromoelectric flux tube model, we have demonstrated that the QN's
will survive against baryon evaporation, if the baryon number of the quark
matter inside the nugget is larger than $10^{42}$. For reasons explained in
\cite{pijush}, this estimate is rather conservative. Sumiyoshi and Kajino
\cite{sumi} have estimated that a QN with an initial baryon number
$\sim 10^{39}$ would survive against baryon evaporation. It should be noted
at this point that the horizon limit on the baryon number at that primordial
epoch is around $10^{49}$. The conclusion from these calculations is
that the larger primordial QN's within the baryon number window
$10^{39-40} \le N_B \le 10^{49}$ are indeed cosmologically stable. (Also
noteworthy in this context is the observation that these numbers are
sufficiently above the limit ($N_B \sim 10^{21}$) which was obtained
by Madsen \cite{madsen} some time ago, below which the QN's could interfere
with the outcome of Big Bang nucleosynthesis.) It is
therefore most relevant to ask what fraction of the dark matter could be
accounted for by the QN's. The central issue that we wish to address in this
letter is whether we can have a scenario where the universe would be closed
with QN's of baryon number within the above range. In other words, can the
proverbial cosmological dark matter, containing 90\% or more of all the
matter in the universe, be made up entirely of such QN's ?

It is well known that in a first order phase transition, the quark and
the hadron phases co-exist. This configuration would be referred to in the
following as the mixed phase. In the quark phase, the universe
consists of leptons, photons as well as the quantum chromodynamic (QCD)
degrees of freedom (massless quarks, anti-quarks and gluons) and is described
by, say, the MIT bag equation of state with an effective degeneracy
$g_Q \approx 51.25$. Note that the baryon number in this phase is carried
entirely by the quarks. The hadronic phase contains baryons, mesons, photons
and leptons and is described by an equation of state corresponding to
massless particles with an effective degeneracy $g_H=17.25$.

The evolution of the universe in the mixed phase at the critical temperature
$T_c$ of the phase transition is governed by the Einstein equation in the
Robertson-Walker spacetime, as described below :
\begin{equation}
\left(\frac{\dot R}{R}\right)^2=\frac{8\pi\epsilon}{3m_{pl}^2}
\end{equation}
\begin{equation}
\frac{d(\epsilon R^3)}{dt}+P\frac{dR^3}{dt}=0
\end{equation}
where $\epsilon$ is the energy density, $P$ the pressure and $G$ the
gravitational constant. Combining  eqs. (1) and (2) with the equation of
state, we determine the scale factor $R(t)$ and the volume fraction of the
quark matter $f(t)$ in the mixed phase as
\begin{equation}
R(t)/R(t_i)=\left[cos\left(arctan\sqrt{3r}-
\sqrt{\frac{3}{r-1}}(t-t_i)/t_c\right)\right]^{2/3}/
\left[cos\left(arctan\sqrt{3r}\right)\right]^{2/3}
\end{equation}
and
\begin{equation}
f(t)=\frac{1}{3(r-1)}\left[tan\lbrace arctan\sqrt{3r}-
\sqrt{\frac{3}{r-1}}\frac{t-t_i}{t_c}\rbrace\right]^2
-\frac{1}{r-1}
\end{equation}
where $r \equiv  g_q/g_h$, $t_c = \sqrt{3m_{pl}^2/8\pi B}$ is the
characteristic time scale for the QCD phase transition in the early universe
and $t_i$ is the time when phase transition starts. 
The bag constant, $B=(245)^4$ MeV$^4$ in our calculations.
           
In the mixed phase, the temperature of the universe remains constant at
$T_c$, the cooling due to expansion being compensated by the liberation of
the latent heat. In the usual picture of bubble nucleation in first order
phase transitions, hadronic matter starts appearing in the quark matter
as individual bubbles. With the progress of time, more and more hadronic
bubbles form, coalesce and eventually percolate to form a network of hadronic
matter which traps the quark phase into finite domains \cite{kodama}. The time
when the percolation takes place is called the percolation time $t_p$,
determined by a critical volume fraction $f_c$, ($f_c \equiv f(t_p)$) of
the quark phase.

Detailed numerical studies on percolating systems yield the result that for
bubbles with the same radial size, $f_c$ is $\sim 0.3$ \cite{iso,stauffer}.
We would also use the same value of $f_c$ here. For the sake of simplicity,
we would also assume that the trapped quark domains are all of the same
size. (In general, this is not true; there ought to be a distribution of
domain sizes. This is an involved problem which we plan to look into in
future. The conclusions of this work, however, are not expected
to be drastically altered.) From eq.(4), we get $t_p - t_i = 0.08 t_c$.
The values of various time scales in the present case are
$t_c=46$~$\mu$sec and $t_p=27$~$\mu$sec.

In an ideal first order phase transition, the fraction of the high
temperature phase decreases from the critical value $f_c$, as these domains
shrink. For the QCD phase transition, however, these domains should become
QN's \cite{witten} and as such, we may assume that the lifetime of the
mixed phase $t_f$ ({\it i.e.}, the time when the cooling due to expansion
starts to dominate again and the temperature of the universe starts falling),
is $\sim t_p$.

The probability of finding a domain of trapped quark matter of
co-ordinate radius $X$ at time $t_p$ is given by \cite{kodama},
\begin{equation}
P(z,x_p)=\exp\left[-\frac{4\pi}{3}v^3t_c^4\int_{x_i}^{x_p}dxI(x)\left(zr(x)
+y(x_p,x)\right)^3\right]
\end{equation}
where $z=X R(t_i)/vt_c$, $x=t/t_c$, $r(x)=R(x)/R(x_i)$ and $I(x)$
is the rate of nucleation per unit volume. $v$ is the radial growth velocity
of the nucleating bubbles, which 
we left as a parameter. $y(x_p,x)$ is given by the
following equation
\begin{equation}
y(x,x\prime)=\int_{x\prime}^x{r(x\prime)}/{r(x\prime\prime)}dx\prime\prime
\end{equation}

Various authors \cite{kodama,cott,landau,kajantie} have proposed different
nucleation rates for the cosmic QCD phase transition. Let us start with the
prescription of ref.~\cite{kodama}, where the nucleation rate is given by the
following expression
\begin{equation}
I(t)=r_T\delta(t-t_i)
\end{equation}
where the prefactor of the thermal nucleation rate $r_T$ is determined
from the requirement $P(0,t_p)=f_c=0.3$, yielding
\begin{equation}
r_T=\frac{3r_c}{4\pi v^3t_c^4}\frac{1}{y(x_p,x_i)}
\end{equation}
with $r_c\sim 1.2$. This leads to
\begin{equation}
P(z,x_p)=\exp\left[-r_c\left(\frac{zf(x_i)}{y(x_p,x_i)}+1\right)^3\right]
\end{equation}
\vskip .2cm
\begin{tabular}{ccc}
~~~~~~~~~~~~~~~~~~~~~~~~~~~~~~~~~~~~~&\psfig{file=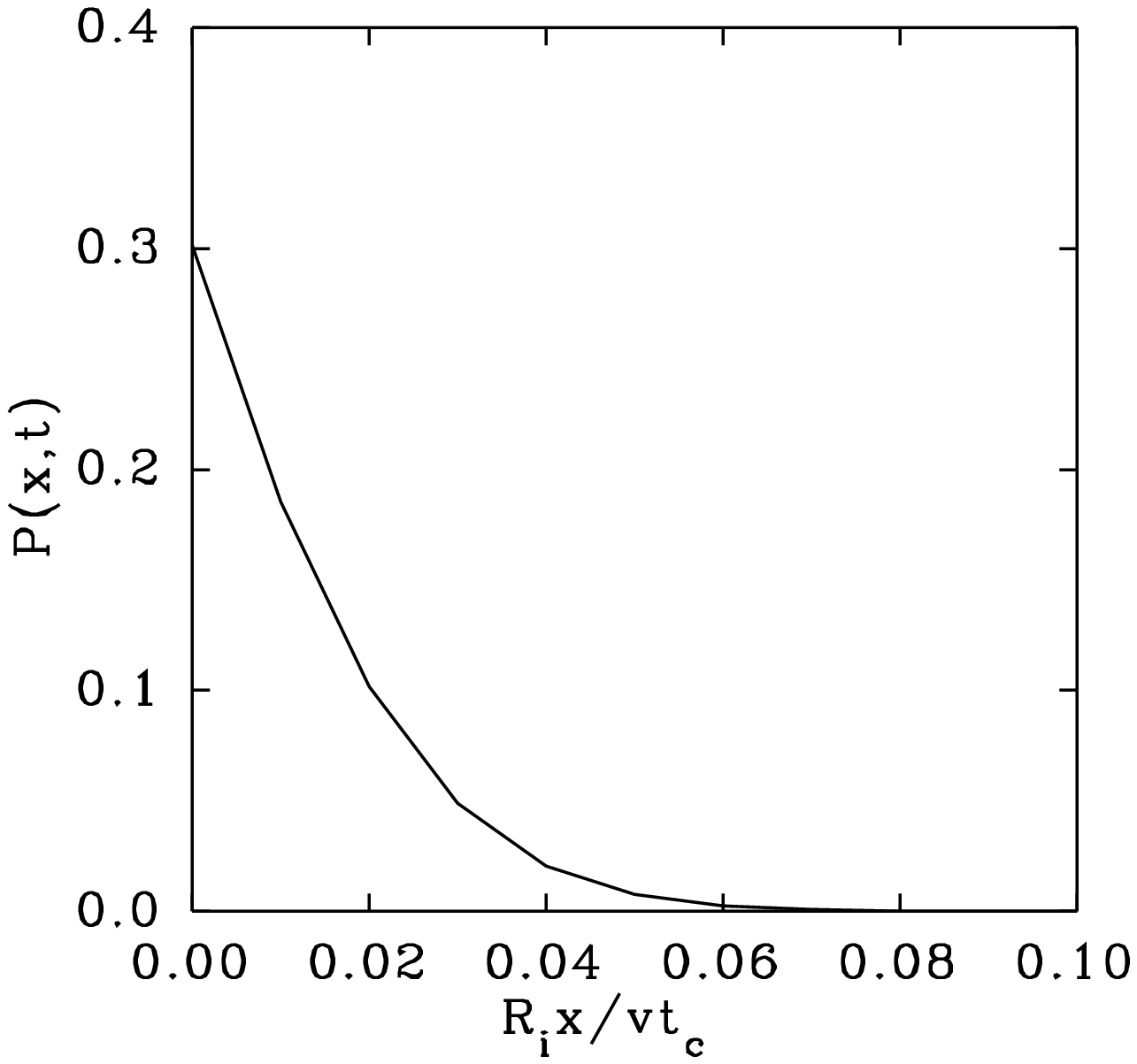,width=2.25in}&\\
\end{tabular}
\vskip .2cm
Fig.1: Probability of finding a domain of co-ordinate size $X$ in the
quark phase at time $t$ as a function of $R_iX/vt_c$ with the nucleation
rate given by eq.(7).
\vskip .2cm
In fig.1, we plot the probability $P(z,x_p)$ as a function of z.
The radius of the trapped quark domain is determined by the length
scale where the  probability falls to $f_c/e$. This implies
\begin{equation}
\frac{r_{QN}}{vt_c} \approx 0.019
\end{equation}

The number density of the QN's $(n_{QN})$ can be obtained by using the
relation $n_{QN}V_{QN}=f_c$ as
\begin{equation}
n_{QN} \approx \frac{1968}{(vt_p)^3}
\end{equation}
In an idealised situation where the universe is closed by the baryonic dark
matter trapped in the QN's, we should have,
\begin{equation}
N_{B}^{H}(t_p)=N_{B}^{QN}n_{QN}V^H(t_p)
\end{equation}
where $N_B^H(t_p)$ is the total number of baryon required to
close the universe $(\Omega_B=1)$ at $t_p$,
$N_B^{QN}$ is the total number
of baryons contained in a single quark nugget and $V_H(t_p)=4\pi(ct_p)^3/3$
is the horizon volume.

Demanding that $v/c\le 1/\sqrt{3}$, we get
\begin{equation}
N_{B}^{QN} \le 10^{-4.7}N_B^{H}(t_p)
\end{equation}

Since the usual baryons constitute only $\sim$ 10\% of the closure density,
a total baryon number of 10$^{50}$ within the horizon at a temperature of
$\sim$ 100 MeV would close the universe baryonically. This would require
$N_B^{QN}$ to be $\le 10^{45.3}$, which is within the survivability limit of
QN's mentioned earlier.

To study the sensitivity of our results to the nucleation rate, we evaluate
the density of QN's for the different nucleation scenarios referred to above.

In the nucleation scenario of Cottingham et al \cite{cott}
the density of QN's is found to be  larger than
the previous case
$n_{QN}=42947/(vt_p)^3$
and the corresponding limit on $N_B^{QN}$ is given by 
$N_{B}^{QN}=10^{-6}N_B^{H}(t_p)$.
For $N_B^H(t_p)\sim 10^{50}$, $N_B^{QN}$ to be $\leq 10^{44}$.
Although we have used the exact nucleation rate obtained in eq.(8)
of \cite{cott},
it is straightforward to show that it
can also be approximated to a delta function.
With the nucleation scenario 
of refs. \cite{landau} and \cite{kajantie},
$N_B^{QN}\leq 10^{44}$ and $10^{44.5}$ respectively 
for the value of the surface tension $=50$ MeV/fm$^2$.

It thus appears that the upper limit on the baryon number of
QN's that would close the universe baryonically is not very sensitive to the
nucleation mechanism ( little about which is confidently known due to
the rudimentary knowledge of the dynamics of the QCD phase transition) and
all available estimates point to the real possibility that stable QN's
could not only be a possible candidate for cold dark matter but they
could even close the universe with total baryon numbers within the
window mentioned earlier.

For the sake of completeness, we would like to mention here that QN's
with baryon number lower than the survivability window would evaporate
rather quickly, leaving a large baryon inhomogeneity. This would still not
pose much problem for the cosmological scenario, as this overdensity
would dissipate, primarily due to neutrino inflation upto temperatures
$\sim$ 1 MeV and then by baryon diffusion, which becomes the dominant
mechanism at lower temperatures \cite{heckler,jedamzik,alam}. Even if the
initial overdensity due to the evaporating QN's could be as high as
10$^{10-12}$, it is found to go down by several orders of magnitude by the
time nucleosynthesis starts. These details would be reported elsewhere
\cite{alam}.

In order for the stable QN's to be a viable candidate for the cosmological
dark matter, they must gravitationally clump, just like normal matter. It
may however be asked what, if any, observational signatures could these
stable QN's have at the present time. To this end, we may estimate the
rate at which such QN's may collide with the earth, in the extreme case
where all these QN's were floating around in the universe. For the various
scenarios \cite{kodama,cott,landau,kajantie} considered here, the rate 
of such collisions (density of QN's$\times$velocity of QN's
($v_{QN}\sim 2\times 10^7$cm/sec)
$\times$surface area of the earth ($\sim 8\times 10^{17}$cm$^2$)) 
turns out to be one in $\sim 10^{15}-10^{16}$ years 
for $v=c/\sqrt{3}$. The collision rate
could be much larger for smaller velocity of the bubble
growth as the rate varies as $1/v^3$.

More realistically, however, the scenarios proposed here would get
substantial experimental support, albeit somewhat indirect, if a sizable
flux of stable strangelets ( heavy nuclear objects with very abnormal
charge-to-mass ratios ) could be detected in the cosmic rays. We may
recall that there has been a report in the literature \cite{japan} that
such objects may indeed form part of the extragalactic cosmic rays.
Extensive plans are under way at this time to look for such objects in
high altitude experiments \cite{madsen2,sibaji}, the outcome of which
would go a long way to shed light on the nature of dark matter.
\vskip .2cm
We are grateful to Dr. Pijushpani Bhattacharjee
for stimulating discussions.

\end{document}